\documentclass[12pt]{article}
 \pdfoutput=1
\usepackage{amssymb}
\usepackage{amsmath}
\usepackage{amstext}
\usepackage{graphicx,epsfig}
\usepackage{epsfig}
\usepackage{verbatim} 
\usepackage{fancybox}
\usepackage{color}
\usepackage{ulem}
\usepackage{enumitem}
\usepackage{subfigure}
\usepackage{bbm}
\usepackage{parskip}
\usepackage{cite}

\newcommand{\Comment}[1]{{}}
\definecolor{MyDarkBlue}{rgb}{0.15,0.15,0.45}
\usepackage[linktocpage=true]{hyperref}
\hypersetup{
colorlinks=true,
citecolor=MyDarkBlue,
linkcolor=MyDarkBlue,
urlcolor=MyDarkBlue,
pdfauthor={Kurt Hinterbichler},
pdftitle={Pseudo-Linear Spin-2 Interactions},
pdfsubject={hep-th}
}

\newcommand{\be}{\begin{equation}}
\newcommand{\ee}{\end{equation}}
\newcommand{\bea}{\begin{eqnarray}}
\newcommand{\eea}{\end{eqnarray}}
\newcommand{\beas}{\begin{eqnarray*}}
\newcommand{\eeas}{\end{eqnarray*}}
\newcommand{\nn}{\nonumber}

\def\({\left(}
\def\){\right)}

\numberwithin{equation}{section}

\begin{document}

\begin{center}
{\LARGE \bf{Pseudo-Linear Spin-2 Interactions}}
\end{center} 

\vspace{2truecm}

\thispagestyle{empty}
\centerline{{\Large James Bonifacio,${}^{\rm }$\footnote{\href{mailto:james.bonifacio@case.edu}{\texttt{james.bonifacio@case.edu}}} Kurt Hinterbichler,${}^{\rm }$\footnote{\href{mailto:kurt.hinterbichler@case.edu}{\texttt{kurt.hinterbichler@case.edu}}} Laura A. Johnson${}^{\rm }$\footnote{\href{mailto:laura.johnson2@case.edu}{\texttt{laura.johnson2@case.edu}}}}}
\vspace{.7cm}

 \centerline{{\it ${}^{\rm }$CERCA, Department of Physics, Case Western Reserve University,}}
 \centerline{{\it  10900 Euclid Ave, Cleveland, OH 44106}} 
 \vspace{.35cm}

\begin{abstract}

We show how the pseudo-linear interactions for spin-2 particles can be straightforwardly extended to include multiple massive and massless fields without introducing ghosts.  When massive spin-2 fields are included, these give simple analogues of bi-gravity and multi-gravity theories with mass terms that break linear gauge symmetries.  The interactions containing only massless particles cannot be written in terms of linearized curvature tensors and are gauge invariant only up to a total derivative. These correspond to known theories that exist in more than four dimensions and evade the no-go results on interacting massless spin-2 particles.

\end{abstract}

\newpage

\section{Introduction}
\parskip=5pt
\normalsize

In four spacetime dimensions, general relativity is the unique ghost-free interacting theory of a single massless spin-2 particle.\footnote{By ghost-free we will mean that the theory is free of Boulware-Deser type ghosts \cite{Boulware:1973my}, meaning the non-linear theory should have the same number of degrees of freedom as the linear theory.  We also implicitly assume that the theory reduces at linear level to the standard linear massless or massive spin-2 theory with the correct sign.}  In higher dimensions this is no longer true and other possibilities arise. These possibilities are classified by whether or not the linearized gauge symmetry is deformed \cite{Wald:1986bj}.  If the linear gauge symmetry is deformed, then this necessarily leads to theories with full general covariance and interactions given by the Lovelock terms \cite{Lovelock:1971yv}.  If instead the linearized gauge symmetry is not deformed, then we can have interactions of the form
\be  \eta^{\mu_1\nu_1\cdots \mu_{2r-1}\nu_{2r-1}} \partial_{\mu_1}\partial_{\nu_1}h_{\mu_2\nu_2}^{} \cdots  \partial_{\mu_{2r-3}}\partial_{\nu_{2r-3}}h_{\mu_{2r-2}\nu_{2r-2}}^{}  h_{\mu_{2r-1}\nu_{2r-1}}^{}, \label{lovelockintroe}\ee
where the spin-2 field is embedded in the symmetric tensor $h_{\mu \nu}$ and the symbol $\eta^{\mu_1\nu_1 \cdots\mu_n\nu_n}$ is defined by products of the flat metric $\eta^{\mu \nu}$ anti-symmetrized over one set of indices,
\be
\label{tensor} 
\eta^{\mu_1\nu_1\mu_2\nu_2\cdots\mu_n\nu_n}\equiv{1\over n!}\sum_p\left(-1\right)^{p}\eta^{\mu_1p(\nu_1)}\eta^{\mu_2p(\nu_2)}\cdots\eta^{\mu_np(\nu_n)} \ ,
\ee
where the sum is over all permutations.
These $r$-th order interactions are non-vanishing in dimensions $D \geq 2 r-1$. They are the lowest-order non-trivial parts of the Lovelock terms when expanding the metric around flat space, $g_{\mu \nu} = \eta_{\mu \nu} + h_{\mu \nu}$, so we refer to them as massless `pseudo-linear' terms.  They are invariant, up to a total derivative, under the linearized gauge symmetry,
\be \delta h_{\mu\nu}=\partial_\mu\xi_\nu+\partial_\nu\xi_\mu,\ee
and are ghost-free by virtue of the epsilon structure and gauge symmetry.

For theories of a single massive spin-2 particle there are also two known types of ghost-free interactions, corresponding to whether the mass term breaks full or linearized diffeomorphisms. The first type are the de Rham-Gabadadze-Tolley (dRGT) interactions \cite{deRham:2010kj} (see \cite{Hinterbichler:2011tt,deRham:2014zqa} for reviews). These are special zero-derivative interactions that can be added to the Einstein-Hilbert or Lovelock terms. Writing these interactions requires introducing a non-dynamical reference metric and results in an explicit breaking of the nonlinear spin-2 gauge symmetry.  The second type are the massive pseudo-linear interactions \cite{Folkerts:2011ev,Hinterbichler:2013eza}. These are ghost-free interactions that can be added to the linearized Einstein-Hilbert term and are massive generalizations of the linearized Lovelock interactions in Eq.~\eqref{lovelockintroe} that break the linear spin-2 gauge symmetry.  These two classes of theories can both be realized as highest strong coupling scale effective field theories for a single massive spin-2 particle~\cite{Bonifacio:2018vzv}.

Now we turn to theories containing multiple spin-2 particles, both massive and massless. The interactions of dRGT massive gravity can be generalized by giving dynamics to the non-dynamical metric. This results in ghost-free bi-gravity, which describes a massive spin-2 particle coupled to the massless spin-2 graviton \cite{Hassan:2011zd} (see \cite{Schmidt-May:2015vnx} for a review).  There are also multi-gravity extensions that describe interactions amongst multiple massive spin-2 particles or interactions of a single massless spin-2 particle with multiple massive spin-2 particles \cite{Hinterbichler:2012cn}. In this case there are restrictions on the couplings and not all choices are fully ghost free \cite{Nomura:2012xr, Scargill:2014wya, deRham:2015cha,Hassan:2018mcw}.  However, in none of these multi-gravity theories are there non-trivial interactions among more than one massless spin-2 particle.

In this paper, we write down explicitly the pseudo-linear analogues of the bi-metric and multi-metric theories.  We will see that, unlike the fully covariant case, there are no restrictions on the structure of the terms and there can be multiple massless fields participating in the interactions. These interactions were suggested in Ref.~\cite{Li:2015fxa}, but here we elaborate on this by explaining the differences for massless and massive particles and by giving the general interactions explicitly.  We also study the Hamiltonian constraints and work out the decoupling limit interactions.  In the special case of multiple massless fields, the allowed ghost-free interactions have been discussed in Ref.~\cite{Bai:2017dwf} and are contained in Ref.~\cite{Chatzistavrakidis:2016dnj}. These massless interactions, which cannot be written as contractions of linearized Riemann tensors, exist only in five or more dimensions and evade the no-go theorem against multiple interacting massless gravitons of Ref.~\cite{Boulanger:2000rq} by having higher-derivative Lagrangians.

\bigskip
\noindent
{\bf Conventions}:
We use the mostly plus metric signature convention, $\eta_{\mu\nu}=(-,+,+,+,\cdots)$. We work in $D$ dimensions with spacetime indices $\mu, \nu, \ldots$ and spatial indices $i, j, \ldots$. Double derivatives are denoted using the condensed notation $\partial_{\mu \nu} \equiv \partial_{\mu} \partial_{\nu}$. 

\section{Multi-field pseudo-linear interactions}

We consider theories with an arbitrary finite number of massless and massive spin-2 fields.
Let there be $n$ different massless fields
\be h_{\mu\nu}^a,\ \ \ a=1,\ldots,n, \, \ee
and $N$ different massive fields
\be H_{\mu\nu}^A,\ \ \ A=1,\ldots,N, \, \ee
with masses $m_A$. Repeated flavor indices $a, b, \ldots$ and $A, B, \ldots$ are implicitly summed over.
The kinetic terms for the massless fields are given by the linearized Einstein-Hilbert action,
\be
\mathcal{L}^{(2)}_{\rm massless}= 3 \delta_{ab}\eta^{\mu_1\nu_1\mu_2\nu_2\mu_3\nu_3} \partial_{\mu_1}h_{\mu_2\nu_2}^a \partial_{\nu_1} h_{\mu_3\nu_3}^b. \ee
These are invariant under $n$ different linear gauge symmetries, 
\be \delta h_{\mu\nu}^a=\partial_\mu \xi_\nu^a +\partial_\nu \xi_\mu^a, \label{multigaugeinve}\ee 
where the $n$ vectors $\xi_\mu^a$ are the generators of the gauge transformations. 
The kinetic terms for the massive fields are given by the standard Fierz-Pauli action \cite{Pauli:1939xp,Fierz:1939ix},
\be
\mathcal{L}^{(2)}_{\rm massive}=3 \delta_{AB}\eta^{\mu_1\nu_1\mu_2\nu_2\mu_3\nu_3} \partial_{\mu_1}\partial_{\nu_2}H_{\mu_2\nu_2}^A H_{\mu_3\nu_3}^B+ m_A^2 \delta_{AB}\eta^{\mu_1\nu_1\mu_2\nu_2}H_{\mu_1\nu_1}^A H_{\mu_2\nu_2}^B. \ee

\subsection{General interactions}
We now systematically list the multi-field generalizations of the pseudo-linear interactions in all dimensions, starting with the cubic interactions, then quartic, etc, and organized by how many massless fields are present. We assume for now that the interactions contain at least one massive field. All interactions containing massless fields should be invariant under the $n$ independent linear gauge symmetries~\eqref{multigaugeinve}. The interaction with $k$ massless fields, $r-k$ massive fields, and $2d$ derivatives, which exists when $D \geq r+d$, is proportional to a constant tensor $T^{(2d)}_{a_1 \ldots a_k A_{k+1} \ldots A_d,A_{d+1}\ldots A_r}$, which carries the coupling constants.  This tensor is symmetric in the indices $a_i$, as are the interactions themselves.  Among the indices  $A_i$, the tensor of coupling constants has symmetry separately amongst two groups of indices, corresponding to groups of differentiated and undifferentiated massive fields (except in the special case of one undifferentiated massive field).  We separate these groups of symmetric indices by a comma. The number of independent couplings is the number of components of these tensors modulo the orthogonal rotations among gravitons with the same mass (since these leave the kinetic terms invariant). 
The explicit interactions are given by the following:
\begin{itemize}

\item cubic terms:
 
\begin{itemize}

\item 0 massless fields:
 \begin{align*}  & T_{A_1 A_2 A_3}^{(0)}\eta^{\mu_1\nu_1\mu_2\nu_2\mu_3\nu_3}H_{\mu_1\nu_1}^{A_1}  H_{\mu_2\nu_2}^{A_2} H_{\mu_3\nu_3}^{A_3},  \\
& T_{A_1,A_2 A_3}^{(2)}\eta^{\mu_1\nu_1\ldots \mu_4\nu_4} \partial_{\mu_1 \nu_1}H_{\mu_2\nu_2}^{A_1}  H_{\mu_3\nu_3}^{A_2} H_{\mu_4\nu_4}^{A_3},  \\
&  T_{A_1 A_2 A_3}^{(4)}\eta^{\mu_1\nu_1\ldots \mu_5\nu_5} \partial_{\mu_1\nu_1}H_{\mu_2\nu_2}^{A_1}  \partial_{\mu_3 \nu_3}H_{\mu_4\nu_4}^{A_2} H_{\mu_5\nu_5}^{A_3}. 
\end{align*}
\item 1 massless field:
\begin{align*} & T_{a_1 A_2 A_3}^{(2)}\eta^{\mu_1\nu_1\ldots \mu_4\nu_4} \partial_{\mu_1 \nu_1}h_{\mu_2\nu_2}^{a_1}  H_{\mu_3\nu_3}^{A_2} H_{\mu_4\nu_4}^{A_3}, \\
& T_{a_1 A_2 A_3}^{(4)}\eta^{\mu_1\nu_1\ldots \mu_5\nu_5} \partial_{\mu_1 \nu_1}h_{\mu_2\nu_2}^{a_1} \partial_{\mu_3 \nu_3}H_{\mu_4\nu_4}^{A_2} H_{\mu_5\nu_5}^{A_3}. 
\end{align*}
\item 2 massless fields:
\[ T_{a_1 a_2 A_3}^{(4)}\eta^{\mu_1\nu_1\ldots \mu_5\nu_5} \partial_{\mu_1 \nu_1}h_{\mu_2\nu_2}^{a_1} \partial_{\mu_3 \nu_3}h_{\mu_4\nu_4}^{a_2} H_{\mu_5\nu_5}^{A_3}.
\]
\end{itemize}

\item quartic terms:
\begin{itemize}
\item 0 massless fields:
\begin{align*} & T_{A_1 A_2 A_3 A_4}^{(0)}\eta^{\mu_1\nu_1\ldots \mu_4\nu_4}H_{\mu_1\nu_1}^{A_1}  H_{\mu_2\nu_2}^{A_2} H_{\mu_3\nu_3}^{A_3} H_{\mu_4\nu_4}^{A_4}, \\
& T_{A_1, A_2 A_3 A_4}^{(2)}\eta^{\mu_1\nu_1\ldots \mu_5\nu_5} \partial_{\mu_1\nu_1}H_{\mu_2\nu_2}^{A_1}  H_{\mu_3\nu_3}^{A_2} H_{\mu_4\nu_4}^{A_3} H_{\mu_5\nu_5}^{A_4}, \\
& T_{A_1 A_2, A_3 A_4}^{(4)}\eta^{\mu_1\nu_1\ldots \mu_6\nu_6} \partial_{\mu_1 \nu_1}H_{\mu_2\nu_2}^{A_1}  \partial_{\mu_3 \nu_3}H_{\mu_4\nu_4}^{A_2} H_{\mu_5\nu_5}^{A_3} H_{\mu_6\nu_6}^{A_4}, \\
& T_{A_1 A_2 A_3 A_4}^{(6)}\eta^{\mu_1\nu_1\ldots \mu_7\nu_7} \partial_{\mu_1 \nu_1}H_{\mu_2\nu_2}^{A_1}  \partial_{\mu_3 \nu_3}H_{\mu_4\nu_4}^{A_2} \partial_{\mu_5 \nu_5} H_{\mu_6\nu_6}^{A_3} H_{\mu_7\nu_7}^{A_4}. 
\end{align*}
\item 1 massless field: 
\begin{align*} & T_{a_1 A_2 A_3 A_4}^{(2)}\eta^{\mu_1\nu_1\ldots \mu_5\nu_5} \partial_{\mu_1 \nu_1}h_{\mu_2\nu_2}^{a_1}  H_{\mu_3\nu_3}^{A_2} H_{\mu_4\nu_4}^{A_3} H_{\mu_5\nu_5}^{A_4}, \\
& T_{a_1 A_2,A_3 A_4}^{(4)}\eta^{\mu_1\nu_1\ldots \mu_6\nu_6} \partial_{\mu_1 \nu_1}h_{\mu_2\nu_2}^{a_1}  \partial_{\mu_3 \nu_3}H_{\mu_4\nu_4}^{A_2} H_{\mu_5\nu_5}^{A_3} H_{\mu_6\nu_6}^{A_4}, \\
& T_{a_1 A_2 A_3 A_4}^{(6)}\eta^{\mu_1\nu_1\ldots \mu_7\nu_7} \partial_{\mu_1 \nu_1}h_{\mu_2\nu_2}^{a_1}  \partial_{\mu_3 \nu_3}H_{\mu_4\nu_4}^{A_2} \partial_{\mu_5 \nu_5} H_{\mu_6\nu_6}^{A_3} H_{\mu_7\nu_7}^{A_4}. 
\end{align*}
\item 2 massless fields:
\begin{align*} & T_{a_1 a_2 A_3 A_4}^{(4)}\eta^{\mu_1\nu_1\ldots \mu_6\nu_6} \partial_{\mu_1 \nu_1}h_{\mu_2\nu_2}^{a_1} \partial_{\mu_3 \nu_3}h_{\mu_4\nu_4}^{a_2} H_{\mu_5\nu_5}^{A_3} H_{\mu_6\nu_6}^{A_4}, \\
& T_{a_1 a_2 A_3 A_4}^{(6)}\eta^{\mu_1\nu_1\ldots \mu_7\nu_7} \partial_{\mu_1 \nu_1}h_{\mu_2\nu_2}^{a_1}  \partial_{\mu_3 \nu_3}h_{\mu_4\nu_4}^{a_2} \partial_{\mu_5 \nu_5} H_{\mu_6\nu_6}^{A_3} H_{\mu_7\nu_7}^{A_4}.
\end{align*}
\item 3 massless fields:
\[ T_{a_1 a_2 a_3 A_4}^{(6)}\eta^{\mu_1\nu_1\ldots \mu_7\nu_7} \partial_{\mu_1 \nu_1}h_{\mu_2\nu_2}^{a_1}  \partial_{\mu_3 \nu_3}h_{\mu_4\nu_4}^{a_2} \partial_{\mu_5 \nu_5} h_{\mu_6\nu_6}^{a_3} H_{\mu_7\nu_7}^{A_4}.
\]
\end{itemize}

\item $\vdots$

\item $r$-th order terms:

\begin{itemize}

\item 0 massless fields:
\begin{align*} & T_{A_1\cdots A_r}^{(0)}\eta^{\mu_1\nu_1\cdots \mu_{r}\nu_{r}} H_{\mu_1\nu_1}^{A_1} \cdots H_{\mu_{r}\nu_{r}}^{A_r}, \\
& T_{A_1,A_2\cdots A_r}^{(2)}\eta^{\mu_1\nu_1\cdots \mu_{r+1}\nu_{r+1}}  \partial_{\mu_1 \nu_1}H_{\mu_2\nu_2}^{A_1}H_{\mu_3\nu_3}^{A_2} \cdots H_{\mu_{r+1}\nu_{r+1}}^{A_r}, \\
& \vdots \\
& T_{A_1\cdots A_r}^{(2r-2)}\eta^{\mu_1\nu_1\cdots \mu_{2r-1}\nu_{2r-1}} \partial_{\mu_1 \nu_1}H_{\mu_2\nu_2}^{A_1} \cdots  \partial_{\mu_{2r-3} \nu_{2r-3}}H_{\mu_{2r-2}\nu_{2r-2}}^{A_{r-1}}  H_{\mu_{2r-1}\nu_{2r-1}}^{A_r}. 
\end{align*}
\item 1 massless field:
\begin{align*} & T_{a_1 A_2\cdots A_r}^{(2)}\eta^{\mu_1\nu_1\cdots \mu_{r+1}\nu_{r+1}}  \partial_{\mu_1 \nu_1}h_{\mu_2\nu_2}^{a_1} H_{\mu_3\nu_3}^{A_2} \cdots H_{\mu_{r+1}\nu_{r+1}}^{A_r}, \\
& T_{a_1 A_2,A_3\cdots A_r}^{(4)}\eta^{\mu_1\nu_1\cdots \mu_{r+2}\nu_{r+2}}  \partial_{\mu_1 \nu_1}h_{\mu_2\nu_2}^{a_1} \partial_{\mu_3 \nu_3}H_{\mu_4\nu_4}^{A_2}  H_{\mu_5\nu_5}^{A_3} \cdots H_{\mu_{r+2}\nu_{r+2}}^{A_r}, \\
& \vdots  \\
&  T_{a_1 A_2\cdots A_r}^{(2r-2)}\eta^{\mu_1\nu_1\cdots \mu_{2r-1}\nu_{2r-1}} \partial_{\mu_1 \nu_1}h_{\mu_2\nu_2}^{a_1} \partial_{\mu_3 \nu_3}H_{\mu_4\nu_4}^{A_2}  \cdots  \partial_{\mu_{2r-3} \nu_{2r-3}}H_{\mu_{2r-2}\nu_{2r-2}}^{A_{r-1}}  H_{\mu_{2r-1}\nu_{2r-1}}^{A_r}.  
\end{align*}
\item 2 massless fields:
\begin{align*} & T_{a_1 a_2A_3\cdots A_r}^{(4)}\eta^{\mu_1\nu_1\cdots \mu_{r+2}\nu_{r+2}}  \partial_{\mu_1 \nu_1}h_{\mu_2\nu_2}^{a_1}  \partial_{\mu_3 \nu_3} h_{\mu_4\nu_4}^{a_2}H_{\mu_5\nu_5}^{A_3} \cdots H_{\mu_{r+2}\nu_{r+2}}^{A_r},  \\
& T_{a_1 a_2A_3,A_4\cdots A_r}^{(6)}\eta^{\mu_1\nu_1\cdots \mu_{r+3}\nu_{r+3}}  \partial_{\mu_1 \nu_1}h_{\mu_2\nu_2}^{a_1}  \partial_{\mu_3 \nu_3} h_{\mu_4\nu_4}^{a_2}\partial_{\mu_5 \nu_5}H_{\mu_6\nu_6}^{A_3} H_{\mu_7\nu_7}^{A_4} \cdots H_{\mu_{r+3}\nu_{r+3}}^{A_r}, \\
& \vdots  \\
&    T_{a_1 a_2 A_3\cdots A_r}^{(2r-2)}\eta^{\mu_1\nu_1\cdots \mu_{2r-1}\nu_{2r-1}} \partial_{\mu_1 \nu_1}h_{\mu_2\nu_2}^{a_1} \partial_{\mu_3 \nu_3}h_{\mu_4\nu_4}^{a_2} \partial_{\mu_5 \nu_5}H_{\mu_6\nu_6}^{A_3}  \cdots    \partial_{\mu_{2r-3} \nu_{2r-3}}H_{\mu_{2r-2}\nu_{2r-2}}^{A_{r-1}}  H_{\mu_{2r-1}\nu_{2r-1}}^{A_r}. 
\end{align*}

\item  $\vdots$ 

\item $r-1$ massless fields:
\[ T_{a_1\cdots a_{r-1}A_r}^{(2r-2)}\eta^{\mu_1\nu_1\cdots \mu_{2r-1}\nu_{2r-1}} \partial_{\mu_1 \nu_1}h_{\mu_2\nu_2}^{a_1} \cdots  \partial_{\mu_{2r-3} \nu_{2r-3}}h_{\mu_{2r-2}\nu_{2r-2}}^{a_{r-1}}  H_{\mu_{2r-1}\nu_{2r-1}}^{A_r}.
\]
\end{itemize}
\item $\vdots$
\end{itemize}

To summarize, the interaction with $2d$ derivatives and $r$ fields, $k$ of them massless, is given by the following expression:
\begin{align} 
  &T^{(2d)}_{a_1 \ldots a_k A_{k+1} \ldots A_d,A_{d+1}\ldots A_r}  \eta^{\mu_1\nu_1\cdots \mu_{r+d} \nu_{r+d}} \partial_{\mu_1 \nu_1}h^{a_1}_{\mu_2 \nu_2}\cdots \partial_{\mu_{2k-1} \nu_{2k-1}} h^{a_k}_{\mu_{2k} \nu_{2k}} \times \nn \\
& \partial_{\mu_{2k+1} \nu_{2k+1}} H^{A_{k+1}}_{\mu_{2k+2} \nu_{2k+2}} \cdots \partial_{\mu_{2d-1}  \nu_{2d-1}} H^{A_d}_{\mu_{2d} \nu_{2d}} H^{A_{d+1}}_{\mu_{2d+1} \nu_{2d+1}}\cdots H^{A_{r}}_{\mu_{r+d} \nu_{r+d}}. \label{eq:general}
\end{align}
We must have $r>d$ otherwise the interaction is a total derivative, and we must have $r+d\leq D$ otherwise it vanishes identically. In particular, the maximum number of particles that can interact at a single vertex is $D$, as in multi-gravity~\cite{Hinterbichler:2012cn}. 

Note that for all the interactions containing massless fields that we have so far considered, the massless fields appear through the linearized Riemann tensor,
\be R_{\mu_1 \mu_2 \nu_1\nu_2 }^{(L)a}\equiv  \partial_{\mu_2}\partial_{ \nu_1} h_{\mu_1 \nu_2}^a-\partial_{\mu_1 }\partial_{ \nu_1} h_{\mu_2 \nu_2}^a +\partial_{\mu_1}\partial_{  \nu_2} h_{\mu_2 \nu_1 }^a-\partial_{\mu_2}\partial_{  \nu_2} h_{\mu_1 \nu_1}^a .\label{linriemanne}\ee
This is apparent because we can make the replacement $\partial_{\mu_2 }\partial_{\nu_2} h_{\mu_1 \nu_1}^a \rightarrow -R_{\mu_1 \mu_2 \nu_1\nu_2 }^{(L)a}/4$ under the anti-symmetric symbol.
The linearized Riemann tensor \eqref{linriemanne} is gauge invariant under \eqref{multigaugeinve} and is the natural field strength for the symmetric potential field $h_{\mu_1 \nu_1}^a$.

\subsection{Massless interactions} \label{ssec:massless}

The interactions involving at least one massive field can be expressed in terms of the field strengths for the massless fields, so they are gauge invariant exactly and not just up to a total derivative.
This is not true of the interactions that involve only the massless spin-2 fields, which generalize the linearized Lovelock interactions \eqref{lovelockintroe}.  These interactions exist in more than four dimensions and contain four or more derivatives. They have the following form:
\begin{itemize}
\item cubic terms:
\be T_{a_1 a_2 a_3}^{(4)}\eta^{\mu_1\nu_1\ldots \mu_5\nu_5} \partial_{\mu_1 \nu_1}h_{\mu_2\nu_2}^{a_1}  \partial_{\mu_3 \nu_3}h_{\mu_4\nu_4}^{a_2} h_{\mu_5\nu_5}^{a_3}. \ee
\item quartic terms:
\be T_{a_1 a_2 a_3 a_4}^{(6)}\eta^{\mu_1\nu_1\ldots \mu_7\nu_7} \partial_{\mu_1 \nu_1}h_{\mu_2\nu_2}^{a_1}  \partial_{\mu_3 \nu_3}h_{\mu_4\nu_4}^{a_2} \partial_{\mu_5 \nu_5}h_{\mu_6\nu_6}^{a_3}  h_{\mu_7\nu_7}^{a_4}. \ee
\item $\vdots$
\item $r$-th order terms:
\be T_{a_1\cdots a_r}^{(2r-2)}\eta^{\mu_1\nu_1\ldots \mu_{2r-1}\nu_{2r-1}} \partial_{\mu_1 \nu_1}h_{\mu_2\nu_2}^{a_1} \cdots  \partial_{\mu_{2r-3} \nu_{2r-3}}h_{\mu_{2r-2}\nu_{2r-2}}^{a_{r-1}}  h_{\mu_{2r-1}\nu_{2r-1}}^{a_r}. \ee
\item $\vdots$
\end{itemize}
We must have $2r-1\leq D$ otherwise these interactions vanish.

These terms are ghost-free interactions among multiple massless spin-2 particles, which have been discussed in Ref.~\cite{Bai:2017dwf} and are contained implicitly in Ref.~\cite{Chatzistavrakidis:2016dnj}.  There are no-go results that seemingly forbid such interactions, but these rely on additional assumptions not satisfied by these terms: either they insist on full general covariance in at least one of the metrics \cite{Aragone:1979bm}, or restrict to two derivative interactions \cite{Boulanger:2000rq}, whereas the interactions here have only linear gauge invariance and are higher-derivative.

Though invariant under the linearized massless gauge symmetries, these terms cannot be written as functions of the linearized curvatures \eqref{linriemanne} because they contain fewer derivatives per field than any function of the field strengths, and thus are only gauge invariant up to a total derivative. These terms are thus distinguished from the more obvious gauge-invariant interactions formed by contracting linearized Riemann tensors and their derivatives~\cite{Wald:1986bj}, which lead to higher-order equations of motion and ghosts. In this sense they are like Chern-Simons terms \cite{Deser:1981wh} for abelian massless spin-2 fields. Other examples of theories with interacting massless spin-2 particles are discussed in Refs.~\cite{Cutler:1986dv, Reuter:1988ig,Boulanger:2000ni}.

\section{Hamiltonian and degrees of freedom}

We now turn to the Hamiltonian structure of the pseudo-linear interactions and argue that they possess the necessary constraints to ensure ghost freedom.

Due to the anti-symmetric structure of the interactions, there can never be more than two timelike indices of fields and derivatives in a single term. This implies that the Lagrangian can be put in the following form after integrating by parts:
\be \mathcal{L}=\mathcal{F}\left(h^a_{ij},\dot{h}^a_{ij},h^a_{0i},H^A_{ij},\dot{H}^A_{ij},H^A_{0i}\right)+ h^b_{00}\mathcal{G}_b\left(h^a_{ij},H^A_{ij}\right)+ H^B_{00}\mathcal{G}_B \left(h^a_{ij},H^A_{ij}\right),
\ee
where $\mathcal{F}$, $\mathcal{G}_b$, and $\mathcal{G}_B$ are functions of the indicated variables.
The momenta conjugate to the spatial tensors $h_{ij}^b$ and $H_{ij}^B$ are then given by
\begin{align}
\pi_b^{ij}\left(h^a_{ij},\dot{h}^a_{ij},h^a_{0i},H^A_{ij},\dot{H}^A_{ij},H^A_{0i}\right) & \equiv \frac{\delta \mathcal{L}}{\delta \dot{h}^b_{ij}}=\frac{\delta \mathcal{F}}{\delta \dot{h}^b_{ij}}, \\ \Pi_B^{ij}\left(h^a_{ij},\dot{h}^a_{ij},h^a_{0i},H^A_{ij},\dot{H}^A_{ij},H^A_{0i}\right) & \equiv \frac{\delta \mathcal{L}}{\delta \dot{H}^B_{ij}}=\frac{\delta \mathcal{F}}{\delta \dot{H}^B_{ij}}.
\end{align}
These can formally be inverted to find the velocities $\dot{h}^b_{ij}$ and $\dot{H}^B_{ij}$ in terms of the phase space variables. The Hamiltonian is then given by
\begin{align}
\mathcal{H} & = \pi_b^{kl}\dot{h}^b_{kl}\left(h^a_{ij},\pi^a_{ij},h^a_{0i},H^A_{ij},\Pi^A_{ij},H^A_{0i}\right)+\Pi_B^{kl}\dot{H}^B_{kl}\left(h^a_{ij},\pi^a_{ij},h^a_{0i},H^A_{ij},\Pi^A_{ij},H^A_{0i}\right) \nn \\
&-\mathcal{F}\left(h^a_{ij},\dot{h}^a_{ij},h^a_{0i},H^A_{ij},\dot{H}^A_{ij},H^A_{0i}\right)- h^b_{00}\mathcal{G}_b\left(h^a_{ij},H^A_{ij}\right)- H^B_{00} \mathcal{G}_B\left(h^a_{ij},H^A_{ij}\right).
\end{align}
The dimension of the phase space spanned by $\{h^a_{ij}, \pi^a_{ij}, H^A_{ij}, \Pi^A_{ij}\}$ is $D(D-1)(N+n)$.

For each massless field $h^a_{\mu\nu}$ there are $D$ first class constraints that generate the gauge transformations \eqref{multigaugeinve}. One of these constraints is
\be
\mathcal{G}_a=0,
\ee 
which is enforced by the Lagrange multiplier $h^a_{00}$.  The other $D-1$ first class constraints are not manifest in the form of Hamiltonian given above, but they are guaranteed to exist by gauge invariance \eqref{multigaugeinve}. They are enforced by Lagrange multipliers that reduce at linear level to the shifts $h^a_{0i}$. 
We do not expect any further first class constraints to be generated, barring any unexpected additional gauge symmetry, so altogether these eliminate $2Dn$ phase space degrees of freedom.

Now consider the constraints coming from the massive fields. The massive shifts $H^{A}_{0i}$ appear algebraically, including a quadratic piece from the mass term, and hence can be integrated out using their equations of motion. The time-like components
$H^A_{00}$ then still appear linearly and enforce one constraint each, 
\be
\mathcal{G}_A =0,
\ee
contributing $N$ primary constraints in total. These constraints are second class since they do not arise from any gauge symmetry. To avoid having half a degree of freedom, these constraints must pair up with each other or with any secondary second class constraints.

With multiple massive fields, there is a danger that some of the primary constraints $\mathcal{G}_A$ do not give rise to secondary constraints and instead form second class pairs with each other.  If this were to happen, there would not be enough constraints to ensure removal of all the Boulware-Deser ghosts.  However, we can argue generally that this will not happen. Since the primary constraints are independent of the conjugate momenta, their Poisson brackets with each other vanish. Moreover, their Poisson brackets with the first class constraints vanish by the definition of first class, so if there were no secondary constraints then we would conclude that $\mathcal{G}_A$ are also first class. This conflicts with our expectation that the massive fields do not generate any additional gauge symmetries, so the primary second class constraints must generate secondary constraints. In fact, there should be $N$ additional second class constraints, since any fewer would result in the matrix of Poisson brackets having additional zero eigenvectors, which would imply the existence of additional first class constraints. The resulting $2N$ second class constraints eliminate $2N$ phase space degrees of freedom.

In total, these constraints imply that the number of physical degrees of freedom is bounded above by
\be \label{eq:dofcount}
\frac{N(D+1)(D-2)}{2} + \frac{n D(D-3)}{2},
\ee  
which is the correct number for $N$ massive and $n$ massless spin-2 particles in $D$ dimensions without any ghosts. This counting holds even if some of the expected secondary constraints do not exist and there are hidden gauge symmetries, since then the corresponding primaries are first class and each remove two phase space degrees of freedom.

\section{Decoupling limit interactions}
We now discuss the decoupling limit interactions of the multi-field pseudo-linear terms. This is a limit that isolates the leading high-energy interactions of the helicity components by taking $m_A \rightarrow 0$ and $M_p \rightarrow \infty$ with the strong coupling scale fixed.  With more than one massive spin-2 field, there is some choice in how to introduce scales, so first we clarify our assumptions and then derive the interactions in the decoupling limit. This pseudo-linear decoupling limit analysis is much simpler than for fully non-linear bi-gravity and multi-gravity, which is discussed in~\cite{Fasiello:2013woa,Noller:2013yja, Noller:2015eda}.

\subsection{Scales and St\"uckelberg fields}
We call the minimum spin-2 mass in our theory $m$ and assume that the ratio $m_A/m$ is independent of the scale $M_p$ for each mass $m_A$. We then consider the decoupling limit that sends all the masses to zero at the same rate.  We also assume that derivatives are suppressed by the mass $m$ and that there is an overall scale $m^2 M_P^{D-2}$, so that a term with $2d$ derivatives has an overall factor of $m^{2-2d} M_p^{D-2}$. This is certainly not necessary and other choices are possible, including different scalings or multiple parametrically distinct $M_p$'s, but this choice is particularly simple. With these assumptions the strong coupling scale is
\be \label{eq:strongscale}
\Lambda_{\frac{D+2}{D-2}} \equiv \left( m^{\frac{4}{D-2}} M_p \right)^{\frac{D-2}{D+2}}.
\ee
This matches the dRGT strong coupling scale and ensures agreement between dRGT and the massive pseudo-linear decoupling limits when there is overlap~\cite{Hinterbichler:2013eza}.

We introduce St\"uckelberg fields by making the linear replacements
\be
H^{A}_{\mu\nu}  \rightarrow \frac{\hat{H}^{A}_{\mu\nu}}{M_p^{D/2-1}}  + \frac{\partial_{\mu} \hat{A}^A_{\nu}+ \partial_{\nu} \hat{A}^A_{\mu}}{m_A M_p^{D/2-1}} + \frac{2 \partial_{\mu} \partial_{\nu} \hat{\phi}^A}{m_A^2 M_p^{D/2-1}}, \quad
h^{a}_{\mu\nu} \rightarrow \frac{\hat{h}^{a}_{\mu\nu}}{M_p^{D/2-1}},
\ee
where the factors of $M_p$ and $m_A$ are chosen to give canonically normalized kinetic terms to the hatted fields.\footnote{Demixing the tensors and scalars at the quadratic level requires an additional linearized conformal transformation, but the decoupling limit can be taken before this demixing.} The Lagrangian is then invariant under $n+N$ separate linear diffeomorphisms and $N$ separate $U(1)$ transformations given by
\be \label{eq:diff}
\delta \hat{h}^a_{\mu \nu} = \partial_{\mu} \xi^a_{\nu}+ \partial_{\nu} \xi^a_{\mu},
\ee 
and
\be \label{eq:stuck}
\delta \hat{H}^A_{\mu \nu} = \partial_{\mu} \Xi^A_{\nu}+ \partial_{\nu} \Xi^A_{\mu}, \quad \delta \hat{A}^A_{\mu} = \partial_{\mu} \lambda^A -m_A \Xi^A_{\mu}, \quad \delta \hat{\phi}^A = - m_A \lambda^A.
\ee 
 With these normalizations the leading terms are those suppressed by the scale $\Lambda_{\frac{D+2}{D-2}}$.

\subsection{Interactions}
We start by looking at interactions containing at least one massive field and comment on the pure massless interactions at the end.  Consider the interaction with $2d$ derivatives and $r$ fields, $k$ of them massless and $r-k$ of them massive.  These are given by \eqref{eq:general} multiplied by $m^{2-2d} M_p^{D-2}$, which schematically takes the form
\be
\sim m^{2-2d} M_p^{D-2} (\partial^2 h)^k (\partial^2 H)^{d-k} H^{r-d}.
\ee
The decoupling limit interactions are then given schematically by
\be
\sim \frac{1}{\Lambda^{(D+2)(r-2)/2}_{\frac{D+2}{D-2}}} (\partial^2 \hat{h})^k (\partial^2 \hat{H})^{d-k} \left(  \hat{H} (\partial^2 \hat{\phi})^{r-d-1} + (\partial \hat{A})^2 (\partial^2 \hat{\phi})^{r-d-2} \right),
\ee
where the undifferentiated massive fields have been replaced either with all scalar terms except for one tensor term or with all scalar terms except for two vector terms. Putting the indices and mass ratios back in, the decoupling limit interactions are given explicitly by
\begin{align}
& \frac{2^{r-d-1}}{\Lambda^{(D+2)(r-2)/2}_{\frac{D+2}{D-2}}} \eta^{\mu_1\nu_1\cdots \mu_{r+d}\nu_{r+d}}  \partial_{\mu_1 \nu_1}\hat{h}^{a_1}_{\mu_2 \nu_2}\cdots \partial_{\mu_{2k-1} \nu_{2k-1}} \hat{h}^{a_k}_{\mu_{2k} \nu_{2k}} \partial_{\mu_{2k+1} \nu_{2k+1}} \hat{H}^{A_{k+1}}_{\mu_{2k+2} \nu_{2k+2}} \cdots \partial_{\mu_{2d-1} \nu_{2d-1}} \hat{H}^{A_{d}}_{\mu_{2d} \nu_{2d}}  \nn \\
& \times  \left( \prod_{l=d+1}^r \frac{m^{2}}{ m^2_{A_l}} \right)  \left( \sum_{i=d+1}^r {m_{A_i}^2\over m^2} {\hat H}^{A_{d+1}}_{\mu_{2d+1} \nu_{2d+1}}\partial_{\mu_{2d+2} \nu_{2d+2}} \hat{\phi}^{A_{d+2}} - \frac{1}{4} \sum_{\substack{i,j=d+1 \\ i\neq j}}^{r} {m_{A_i} m_{A_j}\over m^2} \hat{F}^{A_{d+1}}_{\mu_{2d+1} \mu_{2d+2} } \hat{F}^{A_{d+2}}_{\nu_{2d+1} \nu_{2d+2}} \right) \nn \\
& \times \partial_{\mu_{2d+3} \nu_{2d+3}} \hat{\phi}^{A_{d+3}} \cdots \partial_{\mu_{d+r} \nu_{d+r}}\hat{\phi}^{A_{r}} T^{(2d)}_{a_1 \ldots a_k A_{k+1} \ldots A_d,A_{d+1}\ldots A_r},
\end{align}
where 
$\hat{F}_{\mu \nu}^A \equiv \partial_{\mu} \hat{A}_{\nu}^A-\partial_{\nu} \hat{A}_{\mu}^A$ is the Maxwell field strength and we have used the symmetries of the coefficient tensors to simplify this expression. These interactions are invariant under the massless limit of the gauge transformations in Eqs.~\eqref{eq:diff} and \eqref{eq:stuck}. We can also see that interactions involving particles that are parametrically heavier than $m$ will be suppressed.

The decoupling limit interactions containing vectors only exist for $r>d+1$. When $r=d+1$, the  decoupling limit interactions involve only tensors and have precisely the form of the massless interactions from Sec.~\ref{ssec:massless}. The pure massless interactions also contribute directly to the decoupling limit after normalizing the massless fields. In the decoupling limit of dRGT massive gravity there are no pure tensor interactions, since in this case there are not enough derivatives for such terms to be gauge invariant. 

To demix the tensors and scalars at the quadratic level we need to further redefine
\be
\hat{H}^A_{\mu \nu} \rightarrow \hat{H}^A_{\mu \nu} + {2\over D-2} \eta_{\mu \nu} \hat{\phi}^A.
\ee
This results in additional decoupling limit interactions schematically of the form
\be
\sim \frac{1}{\Lambda^{(D+2)(r-2)/2}_{\frac{D+2}{D-2}}} (\partial^2 \hat{h})^k (\partial^2 \hat{H})^{d-k-l} \hat{\phi} (\partial^2 \hat{\phi})^{r+l-d-1},
\ee
which generalize the Galileon interactions of the  de-mixed dRGT decoupling limit by including couplings to additional spin-2 fields. 

\section{Discussion}

We have discussed the generalization of the pseudo-linear interactions to include multiple massive and massless spin-2 fields. We have written down the interactions explicitly, argued that they are free of the Boulware-Deser ghost by looking at the form of the Hamiltonian, and studied a particular decoupling limit.  Special cases of these interactions have been suggested earlier \cite{Li:2015vwa, Li:2015fxa, Chatzistavrakidis:2016dnj ,Bai:2017dwf}, but as far as we know the multi-field spin-2 case including massive fields has not been systematically studied before. 
These interactions can be thought of as symmetric tensor generalizations of multi-field galileons \cite{Deffayet:2010zh,Padilla:2010de,Hinterbichler:2010xn} or $p$-form galileons \cite{Deffayet:2010zh}, though in the massive case there is no analogue of the galileon symmetry in general.

While the pseudo-linear theories define apparently consistent effective field theories, there are indications that they cannot be embedded in consistent UV-complete theories. For example, terms based on linear gauge invariance have been shown in certain cases to exhibit superluminality, both of fluctuations around non-trivial background solutions \cite{Hertzberg:2016djj,Hertzberg:2017abn} and asymptotically as time advances in the Eikonal phase shift \cite{Camanho:2014apa, Bai:2016hui, Hinterbichler:2017qyt, Bonifacio:2017nnt}. The pseudo-linear theory with a single massive particle is also known to fail $S$-matrix positivity constraints \cite{Bonifacio:2016wcb}, so it cannot admit a standard unitary and analytic Lorentz-invariant completion.

Another limitation of these theories is that the massless spin-2 particles cannot have the standard interactions with matter that mediate long-range forces.  Supposing to the contrary that they did, we could consider the four-point amplitude for the scattering of matter with a massless spin-2 particle. By imposing gauge invariance on this amplitude in the limit in which one of the spin-2 momenta is soft, we would deduce that the massless spin-2 particle couples to itself through the two-derivative Einstein-Hilbert cubic vertex with the same coupling as to matter~\cite{Weinberg:1964ew}. However, this is inconsistent with the linear spin-2 gauge symmetry, since all of the pseudo-linear cubic vertices have at least four derivatives and only fully covariant theories can make use of the two-derivative Einstein-Hilbert vertex.  In fact, this argument implies that the entire massless pseudo-linear spin-2 sector must decouple from Einstein gravity and everything that couples to it. 

Despite these limitations, the multi-field pseudo-linear theories studied in this paper are curious theoretical examples of interacting theories of spin-2 particles. Their existence completes the correspondence between multi-gravity theories and their pseudo-linear analogues. Moreover, we have seen that there are many multi-field pseudo-linear theories without multi-gravity analogues, so they increase the variety of theoretically permitted spin-2 interactions.

\bibliographystyle{utphys}
\addcontentsline{toc}{section}{References}
\bibliography{multipseudolinear-arxiv}

\end{document}